# Kinoform refraction lens with diffraction calculation

L.I.Ognev

*Nuclear fusion institute*
*Russian Research Center «Kurchatov institute»,*
*Moscow, Russia*

## Abstract

By direct numerical simulations of the kinoform refractive lens within the quazioptical approach the effects of shape misalinement were investigated. The effect of $2\pi$ shift compensation was calculated for different orders and imperfections. The performance of kinoform lens was compared with the compound refractive lens and thin lens approximation.

## Introduction

The diffraction model for forming of the x-ray beam passing through refractive lens [1,2] was investigated. In contrast to the paraxial model of the refractive lens with the parabolic refraction profile [3] an arbitrary lens shape can be considered including the teeth-like ("alligator") lens [2]. The diffraction model is based on the integration of the parabolic equation for the complex amplitude of the electric field of the synchrotron radiation $A(x,z)$. The longitudinal axis $Z$ is oriented along the optical axis of the lens. The transversal coordinate is $X$.

$$2ik\,\partial A/\partial z = \Delta_\perp A + k^2[(\varepsilon - \varepsilon_0)/\varepsilon_0]A, \qquad (1)$$

where $k = \sqrt{\varepsilon_0}(\omega/c)$ is wave vector, $\varepsilon$ is dielectric permittivity of the material the lens is made. The initial value of the amplitude before the lens depends on the distance of the radiation source and its size. For coherent source

$$A(x, z = 0) = A_0(x). \qquad (2)$$

In the calculations the Gaussian beam was used

$$A(x, z = 0) = \exp(-x^2/2r_0^2 + i\varphi k x).$$

Where $\varphi$ is input angle to the lens axis. In most calculations was assumed $\varphi=0$. In the calculations the value for real part of the dielectric permittivity is $(1-\varepsilon)=10^{-5}$ and imaginary part is $10^{-7}$ for the x-ray beam energy 12.4 keV. The equation was integrated using step-by-step method [5]. For each step $\delta z$ the "splitting" procedure was applied where the beam propagation between $z$ and $z+\delta z$ was calculated as diffraction in empty space and phase effects for refraction were added afterwards. The error of this procedure was estimated as $O((\delta z)^2)$ [6]. The results for the parabolic lens were compared with geometric optics model. The focus length of the single refractive lens can be obtained as

$$F=2R/(1-\varepsilon), \qquad (3)$$

Where $R$ is radius of curvature of the parabolic lens. For the given parameters of the lens the focus length was estimated as 6625 mm. It can be seen that the results of the equation (1) are in a good agreement with thin lens approximation.

High x-ray absorption in most material limits beam aperture for refractive lens [3]. To overcome this limitation the attempt to withdraw thickness of the length giving $n \cdot 2\pi$ phase shift for given x-ray energy [4]. The most eleborated prototype of the device is kinoform "fern-like" planar lens consisting of many orders of parabolic lenses with subtracted $n \cdot 2\pi$ phase shift thickness. The thickness step for the subtracted layer can be calculated as

$$\delta d = n\lambda \cdot 2/(1-\varepsilon) \qquad (4)$$

In this work only first order kinoform profile was considered. The idea is illustrated with Fig.1, where parabolic shape is splitted in two parts with centaral parabolic zone and outer part with subtracted constant thickness.

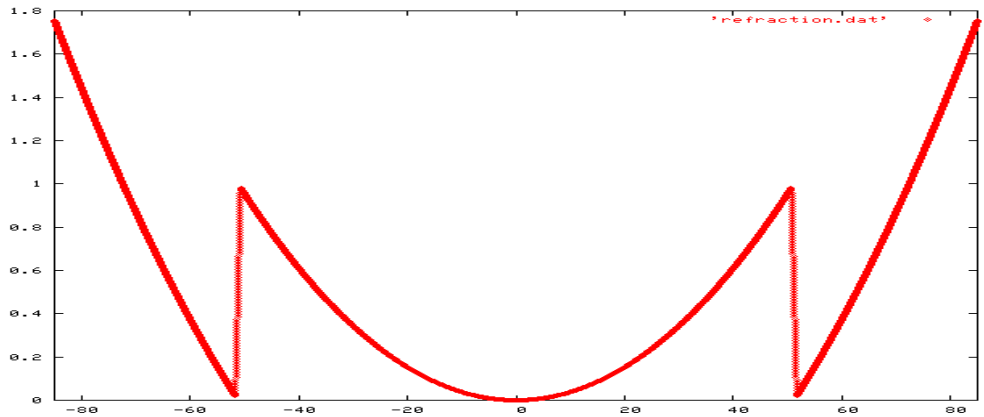

Fig. 1.

The kinoform thickness model of the lens with $n \cdot 2\pi$ phase shift subtraction. The outer parts of the planar lens are parabolas shifted to the bottom and the split between the parts is apodized on the 0.7 μm range. The transversal coordinate is shown in microns.

## Results of calculations

The transmission of x-ray beam through single refractive lens was calculated both for pure parabolic lens and kinoform $4\pi$-phase shift lens and for lenses with reduced thickness by integrating of equation (1). The intensity on the axis behind the parabolic lens depending on the distance is shown on the Fig. 2 with the solid line. The same calculation for the $4\pi$-phase shift kinoform lens with equal curvature of central zone reveals small oscillations of the intensity (lines-points) and higher intensity peak in the focus. The effect may be understood as the result of smaller absorption. The effect is small in this case because the width of the Gaussian beam $r_0$ equals to the width of the central zone. So the portion of the beam transmitting through thick peripherals of the lens was rather small.

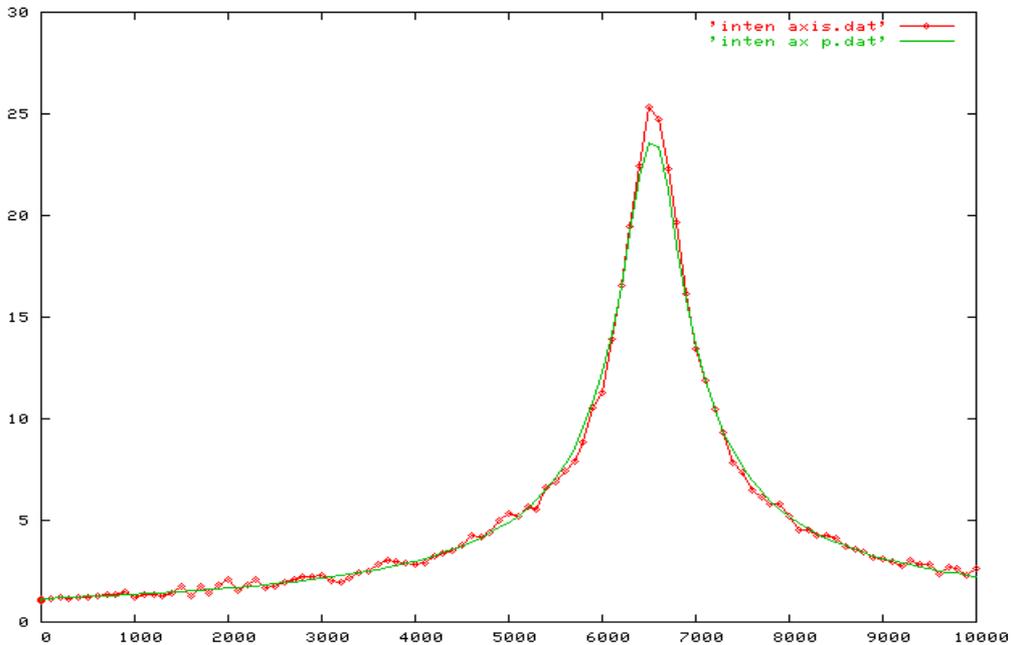

Fig.2
The dependence of the intensity of x-ray beam on the optical axis on the distance behind the refractive lens for coherent 12.4 keV narrow beam for the parabolic lens (solid) and kinoform lens (lines-points) with 40 μm thickness of 0-zone. The distance is measured in millimeters.

The variation of the kinoform lens thickness should effect the intensity of the focused beam behind the lens. The result of the symmetric deviation of the kinoform phase matching from its optimal value was shown on Fig. 3. When the thickness of the lens is reduced by 12.5% from its optimal thickness the intensity is lower by 10% than for the parabolic lens with same curvature. The initial beam width in this case was about the zero-order inner lens width and the initial beam had Gaussian intensity shape. Transversal intensity structure of the focal spot has nearly Gaussian profile.

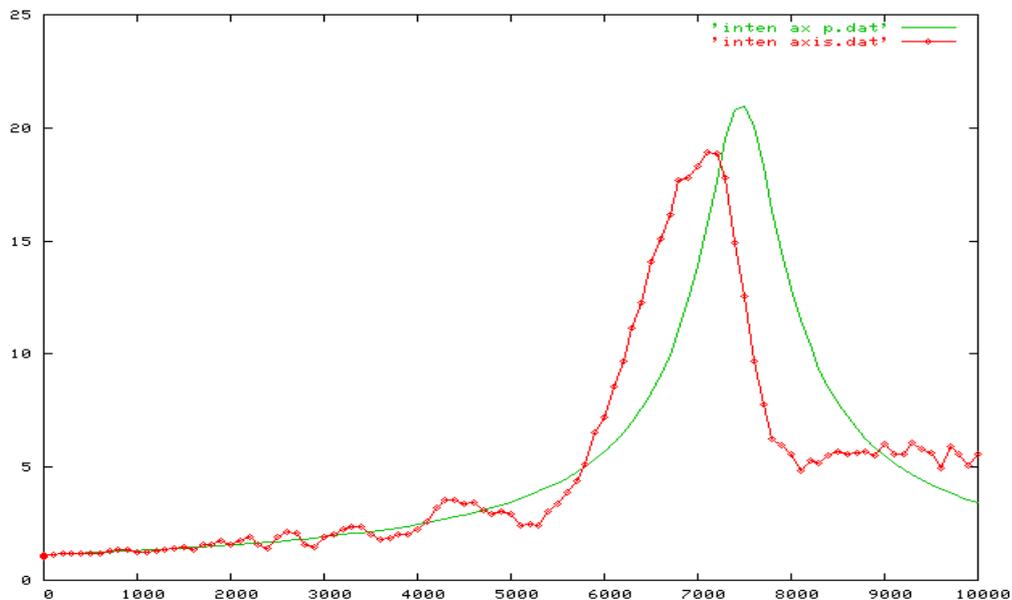

Fig.3

The on axis intensity distribution behind the parabolic lens (solid) and kinoform lens (lines-points) with 35 μm thickness of 0-zone. The distance is measured in millimeters.

The same method was used for investigation of **multiple lens** which is a repetition of many individual lenses separated by empty space as it was made in the 'fern-like' holographic lens [4].

The near field intensity distribution behind the lens is shown on Fig. 4. Both the increased absorption and the phase gap at the edge of the central refraction zone result in intensity fringes behind the lens. The intensity fringes are shown on Fig. 4. In the bottom of the figure the effective thickness of the lens in the vicinity of the gap is shown for comparison. The drop of the intensity in the vicinity of the kinoform gap is due to absorption of the beam. The oscillations look like interference pattern after the sharp edge.

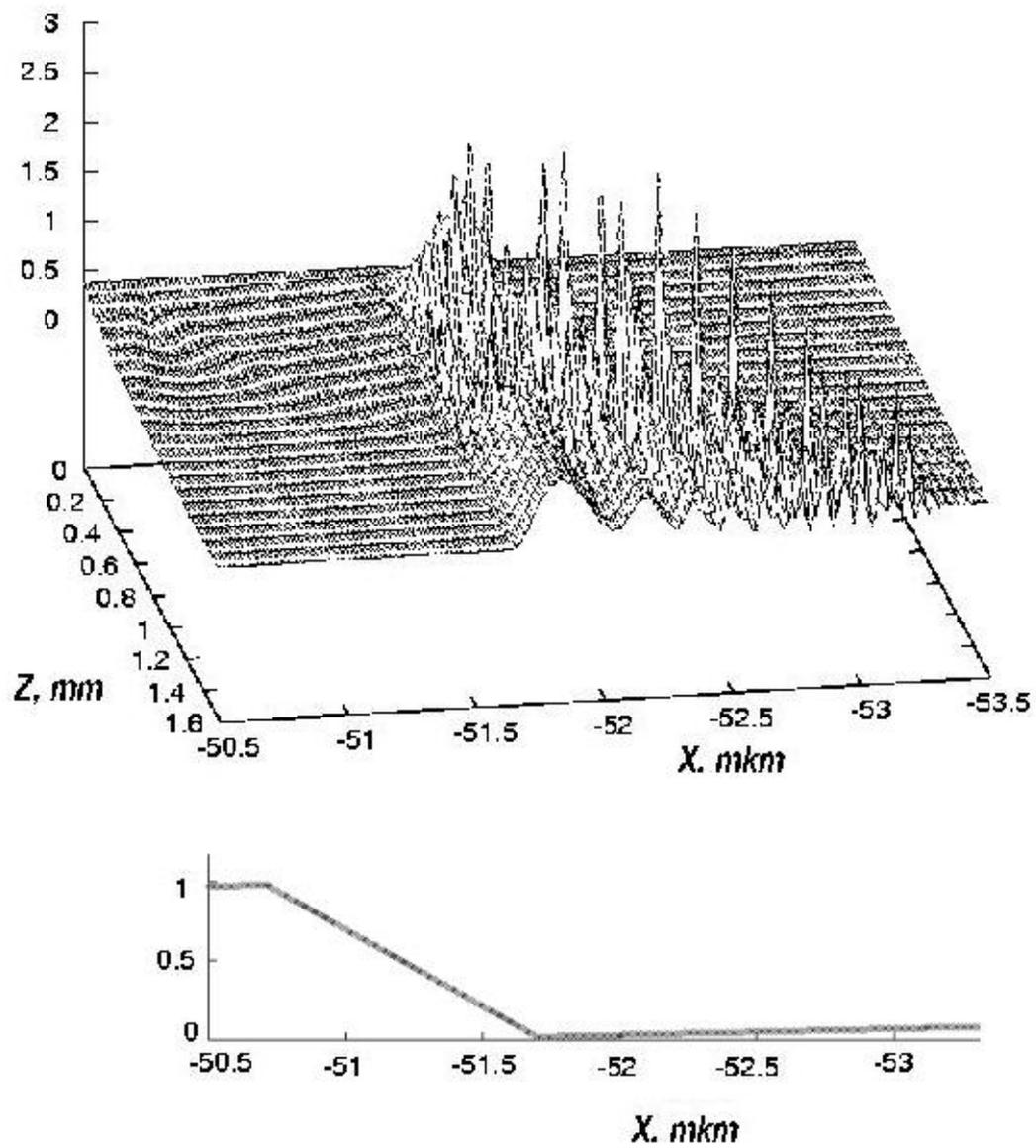

Fig.4

Intensity distribution behind the single kinoform lens. In the bottom of the figure the effective thickness of the refractive lens lens in the visinity of the thickness gap is shown as the check point.

The effect of symmetrical phase mismatch of individual lenses on multiple kinoform lens

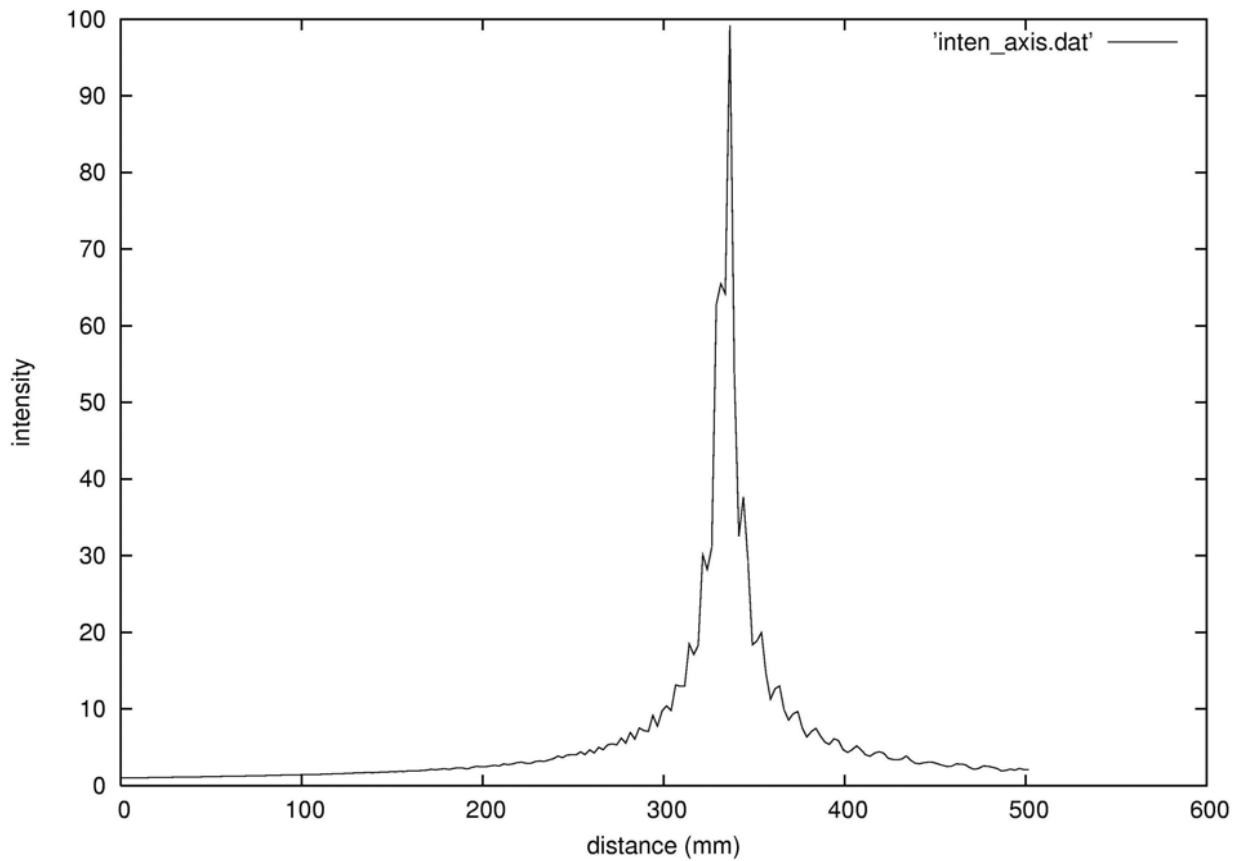

Fig. 5

The on-axis intensity behind multiple kinoform refractive lens
containing 20 components with symmetrically distorted phase
matching. The distance between simple lenses was 40 μm.

was investigated both in the near field intensity and far field on axis intensity near the focal spot. The individual lenses had reduced by 7% thickness. The near field fringes after the lens are similar to the case shown on Fig. 4 before. But the on-axis intensity behind multiple kinoform refractive lens containing 20 components with symmetrically distorted phase matching reveal pronounced interference oscillations as shown on Fig.5. Distance between simple lenses was 40 μm as in the optimal case. The stucture of the focal spot is shown on Fig. 6. No strong spacial peculiarities can be noticed in the case of symmetric distortions.

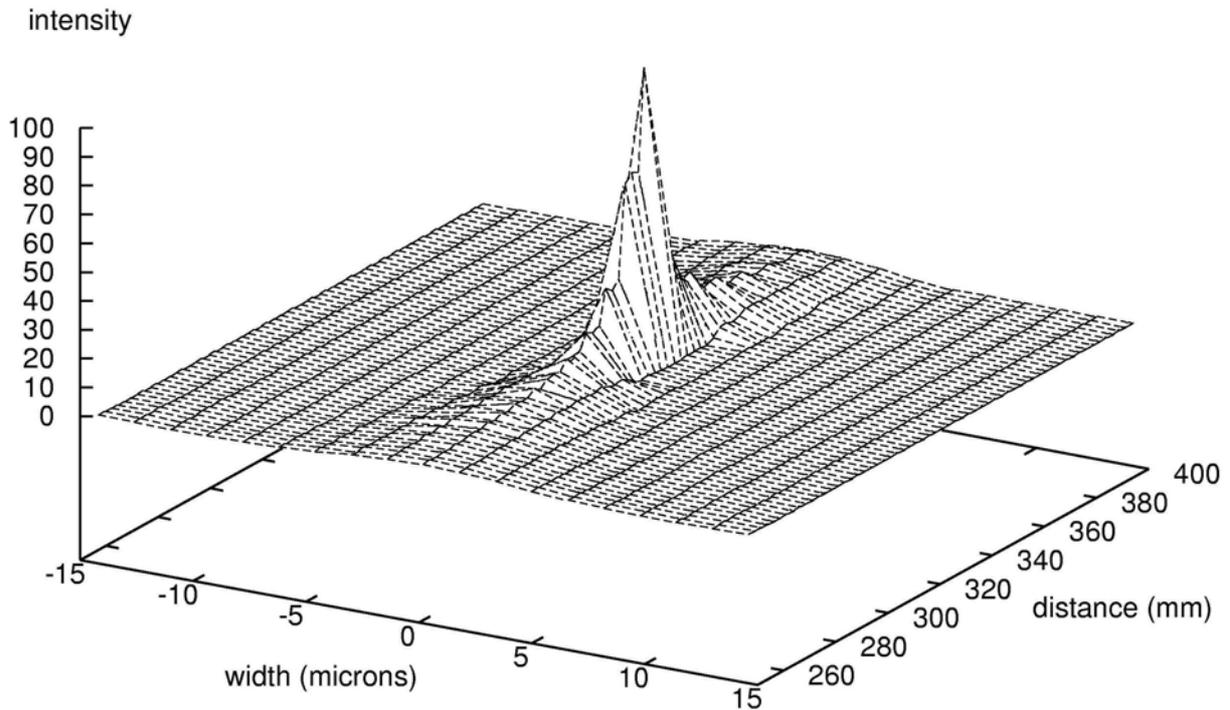

Fig. 6

Focal spot structure behind multiple kinoform refractive lens containing 20 components with distorted phase matching. Distance between simple lenses was 40 μm.

## **Conclusion**

The diffraction model of the kinoform refractive x-ray lens shows that sharp edge at the border between 0-order and 1-order zones gives rise to near field intensity fringes. The effect can be experimentally observed. But in the far field zone there no drastic effect and focal spot has simple structure. The thickness symmetric variation of the kinoform lens results in intensity oscillations on the optical axis and reduction of the intensity in the focal spot compared to the parabolic lens. But no strong variations of the focal structure were found. The model can be extended to statistical variation of individual lenses shape and to partially coherent x-ray beams.

# References

1. A. Snigirev, V. Kohn, I. Snigireva, B. Leneler, A compound refractive lens for focusing high-energy X-ray, Nature, v.384, n.7, p.49-51 (1996).

2. B. Cedestrom, R. Cahn, M. Danielsson, M. Lundquist, D. R. Nygren, Focusing hard X-rays with old LPs, Nature, v.404, 951 (2000).

3. V. G. Kohn, On the Theory of X-ray Refractive Optics: Exact Solution for a Parabolic Medium, JETP Letters, v.76, n.10, pp. 600-603 (2002).

4. I. Snigireva, A. Snigirev, C. Rau, T. Weitkamp, V. Aristov, M. Grigoriev, S. Kuznetsov, L. Shabelnikov, V. Yunkin, M. Hoffman, E. Voges, Holographic x-ray optical elements: transition between refraction and diffraction, NIM (A), v.A467-468, p.982-985 (2001).

5. T. A. Bobrova, L. I. Ognev "Numerical simulation of gliding reflection of X-ray beam from rough surface", Preprint of Kurchatov Institute, IAE-6051/11, Moscow, 1997. http://arXiv.org/abs/physics/9807033.

6. J. A. Fleck Jr., J.R.Morris, M.D. Feit, Time-dependent propagation of high energy laser beams through the atmosphere, Applied Physics, v.10, n.2, p.129-160 (1976).